\documentstyle[12pt]{article}
\def\np    { Nucl. Phys. }

\def\pl    { Phys. Lett. }

\def\beqa{\begin{eqnarray}}
\def\eeqa{\end{eqnarray}}
\def\parn              {  \par\noindent }
\def\parbigskip        {  \par\bigskip  }

\def\parbigskipn        {  \par\bigskip\noindent  }

\def\papertitlepage{\baselineskip 3.5ex \thispagestyle{empty}}
\def\Title#1{\vspace{1.5cm}\begin{center}
 {\Large\bf #1} \end{center} 
\vspace{1cm}}
\def\Authors#1{\begin{center} {\large\it #1} \end{center}}
\def\Abstract{\vspace{0.3cm}\begin{center} {\large\bf Abstract} 
           \end{center} \parbigskip}
\def\ICRRnumber#1#2#3{\hfill \begin{minipage}{3cm} #1
              \parn #2 \parn #3 \end{minipage}}
\begin{document}
\papertitlepage
\vspace*{-1 cm}
\ICRRnumber{ }{}{ }
\Title{ The mass formula for a fundamental string\\
\vskip 1.5ex
as a BPS solution of a D-brane's worldvolume
} 
\Authors{{\sc\  Takeshi Sato
\footnote{tsato@icrr.u-tokyo.ac.jp}} \\
 \vskip 3ex
 Institute for Cosmic Ray Research, University of Tokyo, \\
5-1-5 Kashiwa-no-ha, Kashiwa, Chiba, 
277-8582 Japan \\
}
\Abstract

We propose
a (generalized) ``mass formula'' for 
a fundamental string described as 
a BPS solution of a D-brane's worldvolume.
The mass formula is obtained by using the Hamiltonian density
on the worldvolume, based on 
transformation properties
required for it.
Its validity is confirmed 
by investigating the cases of 
point charge solutions of 
D-branes in a D-8-brane (i.e. curved) background,
where the mass of each of the corresponding strings is 
proportional to the geodesic distance
from the D-brane to the point 
parametrized by the (regularized) value 
of a transverse scalar field.
It is also shown that
the mass of the string agrees with
the energy defined on the D-brane's worldvolume
only in the flat background limit,
but the agreement does not always hold 
when the background is curved.

\newpage

\section{Introduction}

To gain better understanding of string 
theories and M-theory,
intersecting branes 
have played an important role, and
worldvolume analyses have been powerful approaches to investigate
the intersecting 
branes\cite{cal1}\cite{how3}\cite{gib1}\cite{tow6}\cite{gaun1}. 
In 
ref.\cite{cal1}\cite{how3}\cite{gib1}
solutions of
worldvolume field theories of branes 
(in flat backgrounds) 
with nontrivial 
worldvolume gauge fields were obtained.
In the case of D-branes,
it was shown 
that 
an appropriate excitation of one of
the transverse scalar fields is needed
in order to obtain a
{\it supersymmetric} (i.e. BPS)
point charge solution of the worldvolume gauge field\cite{cal1}.
Each of the solutions was interpreted as 
a fundamental string ending on the D-brane 
on the basis of the fact that 
the (regularized) energy of 
each solution defined on the worldvolume
is proportional to the 
(regularized) value of the scalar field,
which is considered to be the length of
the string.\footnote{Since the energies of the point charge solutions
are infinite, 
(UV) regularization is needed to calculate them.}
This interpretation
is also consistent with the charge conservation suggested in 
ref.\cite{openpbrane} (see also ref.\cite{surgery}).
These analyses are very important in that
they made clear 
``how one of intersecting two branes is described''  
from the viewpoint of another brane's worldvolume. 

In fact, however, 
{\it the energy 
defined on the brane's worldvolume 
does not always agree with 
the target-space mass of the string}, 
though the agreement of the two holds for the case of ref.\cite{cal1}.
We can easily understand this by considering
the fact that
the worldvolume energy depends on the definition of the time
of the worldvolume, 
while the mass of the string should not depend on it.
That is,
the derivation of the mass of the string
must be considered more carefully.
The main purpose of this paper is 
to present 
a generalized ``mass formula'' of a fundamental string 
described as a BPS solution of a D-brane's worldvolume,
which holds independent of definition of its worldvolume time.

Our idea to obtain the mass formula 
is as follows: 
the mass of the string does not always agree with
the worldvolume energy, to be sure, but 
it is also true
that the two are very close to each other, since 
the two give the same results
at least in the cases discussed in ref.\cite{cal1}.
So, we construct
the mass formula in a heuristic way,
by using the Hamiltonian density defined on the worldvolume,
based on the invariance
of the mass under the
coordinate transformations of the worldvolume.
Moreover, 
we also construct explicitly
point charge solutions 
(with appropriate excitations of single transverse 
scalar fields) 
of branes' worldvolumes in a {\it curved} background.\footnote{Some 
worldvolume solutions of branes in a curved (brane) backgrounds 
are discussed for other purposes in 
ref.\cite{sjrey}\cite{sato3}.}
There are two advantages to consider these solutions:
First, in each of the cases, the time component of 
the induced worldvolume metric $\tilde{g}_{00}$  
(as well as spatial ones) becomes nontrivial,
So, the differences
originating from the contribution of $\tilde{g}_{00}$ become
apparant
in discussing the mass formula.
Second, 
in each case of the solutions, 
{\it the mass of
the corresponding fundamental string should be 
proportional to
the geodesic distance from the brane to the point 
parameterized by the (regularized) 
value of the scalar field}, 
and that the proportional coefficient should be 
an appropriate tension of the string, 
as discussed in ref.\cite{sen1}. 
This requirement is very tight.
So, once we find out
the quantity which gives the mass stated above,
it is expected to be the correct mass formula,
even if it is constructed by hand.
We will construct it based on this idea.
(The discussion on the worldvolume interpretation of the string mass 
will be given finally in section 3.)

The worldvolume theories we discuss here are
the two cases: those of 
a test D-4-brane and a test D-8-brane both 
embedded parallel to (a subspace of)
the worldvolume of
the D-8-brane background\cite{pol1}\cite{berg3}
(i.e. a massive IIA background).
First, we present the two reasons to choose this background:
One is that
this background has only one transverse coordinate,
leading to the fact that 
the harmonic function depends linearly on the coordinate.  
Only in this case,
we can obtain {\it explicitly} 
the exact solutions of the worldvolumes 
without any extra assumption
(as we will see later).
We note that obtaining the explicit form is
is crucial not only in order to find out the mass formula {\it by hand}
and but for other discussions. 
Another reason is that
choosing this background,
we can see the supersymmetry preserved
in the solutions
by using superalgebras in a massive IIA background via brane probes''
in ref.\cite{sato2}.
\footnote{In the previous paper\cite{sato3}
we have confirmed that  
the supersymmetry preserved in the solution of brane's 
worldvolume can be derived via ``superalgebras in brane
backgrounds''
(see also \cite{sato1}).}
From the supersymmetry, we can confirm 
that the solutions are BPS states
and that their target-space interpretation
is consistent.
Next, we explain why we choose the two worldvolume theories
embedded in those ways:
This is because
at least one overall transverse space is needed
{\it after} embedding (test) D-branes' worldvolumes
in order to obtain supersymmetric point charge solutions.
In the case of 
a D-2-brane (a D-6-brane),
only the intersection with the background D-8-brane 
on a string (a 5-brane) leads 
to the preservation of supersymmetry,\footnote{
We do not discuss bound states like (6$|$D6,D8)
since 
the D-8-brane solution is ``singular'' just on the D-8-brane 
hyper-surface.}
but, 
there is no overall transverse space. 
A D-0-brane
is not adequate to this worldvolume description
since there is no world space.
So, we consider 
the above two cases,
which preserve at this moment 1/4 and 1/2
supersymmetry, respectively.

Concrete procedures are in the following:
In each of the two cases we construct 
explicitly an point charge solution 
with an appropriate excitation of
the only overall transverse scalar field.
The consistency of 
the interpretation of each solution as a fundamental string
is confirmed in two ways:
by discussing its behavior 
in the flat background limit
and by checking its preserved supersymmetry.
Then, we propose the generalized mass formula,
and examine
whether it gives the geodesic distance
multiplied by an appropriate tension of the string.
Moreover, we also discuss
the condition that the mass of the string agrees with
the energy defined on the D-brane's worldvolume.

The organization of this paper is as follows:
In section 2 we construct the point charge solutions 
and discuss their mass formula.
In section 3 we give the conclusion and some discussions
on the consistency and the interpretation
of the mass formula

The notations in this paper is as follows:
We use ``mostly plus'' metrics for both spacetime and
worldvolumes.
We denote coordinates of each p-brane's worldvolume as 
$\xi^{i},\xi^{j},..$ (i,j =0,1,..,p), 
those of 10D spacetime as $x^{m}, x^{n},..$ (m,n=0,..,9), 
fermionic coordinates as $\theta^{\alpha},\theta^{\beta},..$,
and those of superspace 
as $Z^{M}$. 
We use {\it hatted letters} 
($\hat{M},\hat{m},\hat{\alpha}..$) 
for {\it all the local Lorentz frame indices}
and  {\it under-barred} letters ($\underline{m}, 
\underline{i}$) for {\it spatial} indices 
(but not time one), respectively.
We denote gamma matrices as
$\Gamma_{\hat{m}}$, which are all real and satisfy 
$ \{ \Gamma_{\hat{m}} , \Gamma_{\hat{n}} \} 
= 2\eta_{\hat{m}\hat{n}}$. 
$\Gamma_{\hat{0}}$ is antisymmetric and others 
symmetric. Charge Conjugation is ${\cal C}=\Gamma^{\hat{0}}$.

\section{Point charge solutions 
of 10D IIA D-branes' worldvolumes and their mass formula
}
\setcounter{footnote}{0} 
In this section we construct the point charge solutions corresponding
to fudamental strings and discuss their mass formula.

The D-p-brane action
in a general 10D massive 
IIA background\cite{rom1}\cite{berg3} 
takes the form\cite{green1}\cite{berg5}
\beqa
S_{{\rm Dp}} = S_{{\rm Dp}}^{BI}+S_{{\rm Dp}}^{WZ}
&=& -T_{p}\int d^{p+1}\xi e^{-\phi}
\sqrt{-{\rm det}(\tilde{g}_{ij}+{\cal F}_{ij})}\nonumber\\
& &+T_{p}\int [\tilde{C} e^{{\cal F}}|_{{\rm (p+1)-form}}
+\frac{m}{(p/2+1)!}V(dV)^{p/2}],
\label{Dpaction}
\eeqa
where $\tilde{g}_{ij}
=E_{i}^{\ \hat{m}}E_{j}^{\ \hat{n}}\eta_{\hat{m}\hat{n}}$
is
the induced worldvolume metric
where $E_{i}^{\ \hat{m}}=\partial_{i}Z^{M}
E_{M}^{\ \hat{m}}$ where $E_{M}^{\ \hat{N}}$ is the supervielbein. 
$\phi$ is the dilaton field and
${\cal F}_{ij}$ are the components of a modified 
worldvolume 2-form field strength
\beqa
{\cal F}=dV-\tilde{B}_{2}
\eeqa
of the worldvolume 1-form gauge field $V$ where 
$\tilde{B}_{2}$ is the worldvolume 2-form induced by the superspace
NS-NS 2-form gauge potential $B_{2}$.
$\tilde{C}$ is a formal sum of worldvolume r-forms 
$\tilde{C}^{(r)}$ (r=1,3,5,7,9)
induced by the superspace R-R r-form gauge potentials $C^{(r)}$.
$m$ is a mass parameter 
which is the dual of the 10-form 
field strength $F^{(10)}$
of a R-R 9-form $C^{(9)}$\cite{pol2}\cite{berg3}.
$T_{p}$ is the ``formal'' (but not physical) tension of the D-p-brane
which is given by\cite{green1}\footnote{
We choose $\alpha'=1$ in this paper.}
\beqa
T_{p}=\frac{1}{(2\pi)^{p}}.
\eeqa 

We take the background of the action (\ref{Dpaction})
to be the D-8-brane solution 
given by\cite{pol1}\cite{berg3}
\beqa
ds^{2} &=& H^{\epsilon/2}dx^{\mu}dx^{\nu}\eta_{\mu\nu}
+H^{-5\epsilon /2-2}dy^{2}
\nonumber\\
e^{\phi} &=& H^{5\epsilon /4}\nonumber\\
C^{(9)}_{01\cdots 8}&=&H^{\epsilon}\label{d8sol}
\eeqa
where $x^{\mu}$ and $x^{\nu}$ ($\mu,\nu$=0,..,8 )
are the spacetime coordinates parallel to the D-8-brane and
$y$ is a single transverse 
coordinate. $\epsilon$ is 
a nonzero parameter which cannot be determined by the equations
of motions of 10D massive IIA supergravity.
We note that the solution
(\ref{d8sol}) with $\epsilon=-1$ is the standard form of
the D-8-brane solution
since it is obtained via T-duality
from the other D-p-brane solutions\cite{berg3}.
$H=H(y)$ is a harmonic function on $y$.
In this paper we set 
\beqa 
H(y)=c_{1}+\frac{m}{|\epsilon|}|y|,\label{harm1}
\eeqa
which means that the D-8-brane lies at $y=0$.
We choose $c_{1}>0$ and $ m>0$ to avoid a singularity at
$y=0$ and to get a real dilaton.
We note that
the solution (\ref{d8sol}) becomes the flat spacetime metric
in the massless limit 
\beqa
\left\{
  \begin{array}{lll} 
m &\rightarrow & 0 \\  
c_{1} &\rightarrow &1 \ ({\rm \ via\  diffeomorphism}).
\end{array}
\right. \label{masslesslim}
\eeqa
The Killing spinor of (\ref{d8sol}) has the form 
$\varepsilon = H^{\epsilon/8}\varepsilon_{0} $ where $\varepsilon_{0}$
has a definite chirality,
i.e. $ \Gamma_{\hat{y}}
\varepsilon_{0}=+\varepsilon_{0}$ for $y>0$ and $\Gamma_{\hat{y}}
\varepsilon_{0}=-\varepsilon_{0}$ for $y<0$.

We first consider 
a point charge solution 
of a D-4-brane worldvolume
parallel to the background D-8-brane.
Since the solutions we construct here is a bosonic one, 
we set fermionic coordinates $\theta$ to be zero.
Moreover, we consider the ansatz
\beqa
\left\{
  \begin{array}{ll} 
    x^{i}=\xi^{i} 
      & (i=0,1,\cdots,4 )\\
    x^{5},x^{6},x^{7},x^{8}:&{\rm  \ \ constants}\\
    y=y(r)>0& \ \\
    V_{0}= V_{0}(r),& V_{\underline{i}}=0\\
  \end{array}
\right.\label{pointsolansatz}
\eeqa
where $r$ is defined as 
$r\equiv\sqrt{\Sigma_{\underline{i}=1}^{4}
(\xi^{\underline{i}})^{2}}$.
We note that 
the upper two columns of (\ref{pointsolansatz}) mean that
the D-4-brane is embedded in the 1234-hyper-plane.
That is, from target-space point of view,
we consider
the following intersection of three branes: 
\begin{center}
\begin{tabular}{lllllcrrrrr}
{\rm background D8 (at $y=0$)}:& 0&1&2&3&4&5&6&7&8&-\\
{\rm worldvolume D4 (at $y=y_{0}$)}:& 0&1&2&3&4&-&-&-&-&-\\
{\rm fundamental string } :& 0&-&-&-&-&-&-&-&-&9
\end{tabular}
\end{center}
where $y_{0}$ is a positive constant (and $x^{9}=y$). 

Then, since $V_{\underline{i}} =\theta=0$,
$ S_{{\rm D4}}^{WZ}$ does not contribute to the equations of motion,
and
the equations of motion to solve
are given by
\beqa
\frac{\delta{\cal L}^{BI}}{\delta x^{m}}=
\partial_{i}[\frac{\delta{\cal L}^{BI}}{\delta \partial_{i}x^{m}}],
\ \ 
\frac{\delta{\cal L}^{BI}}{\delta V_{i}}&=&
\partial_{j}[\frac{\delta{\cal L}^{BI}}
{\delta \partial_{j}V_{i}}].\label{eom1}
\eeqa
So, let us examine more about ${\cal L}^{BI}$
The induced worldvolume metric 
$\tilde{g}_{ij}$ is given by
\beqa
\tilde{g}_{ij}=
\left(
\begin{array}{cc}
-H^{\epsilon/2}&0\\
0&H^{\epsilon/2}\cdot [\delta_{\underline{i}\underline{j}}
+H^{-3\epsilon-2}
\partial_{\underline{i}} y\partial_{\underline{j}} y]
\end{array}
\right), \label{wvmetric1}
\eeqa
whose determinant is 
${\rm det}\tilde{g}_{ij}=H^{5\epsilon/2}
[1+H^{-3\epsilon-2}(\partial 
y)^{2}]$.
At this moment,
det
$(\tilde{g}_{ij}+{\cal F}_{ij})$ arising in  ${\cal L}^{BI}$
is very complicated. 
However, setting the condition
\beqa
\partial_{\underline{i}}y=\partial_{\underline{i}}V_{0},
\label{yveual1}
\eeqa
results in
the simple form of the determinant:
${\rm det}(\tilde{g}_{ij}+{\cal F}_{ij})
=H^{5\epsilon/2}$.\footnote{
It is shown that 
no supersymmetry is preserved in this case without (\ref{yveual1}),
by using a superalgebra via brane probe
appearing later (eq.(\ref{spalgd4})).
So, (\ref{yveual1}) is considered to
correspond to the BPS condition.
}
Then, the equations of motion 
(\ref{eom1}) become 
the following two simple equations
\beqa
\sum_{\underline{i}}
\partial_{\underline{i}}(H^{-3\epsilon-2}\partial_{\underline{i}}y)
&=&0\label{eq11}\\
\sum_{\underline{i}}
\partial_{\underline{i}}(H^{-2\epsilon-1}\partial_{\underline{i}}y)
&=&0.\label{eom2}
\eeqa
((\ref{eq11}) arises from the 9th component of
the former of (\ref{eom1}), (\ref{eom2}) from the time
component of the latter of (\ref{eom1}), and the others are solved.)
So, requiring the two equations to be 
compatible,
it needs to hold $\epsilon=-1$, 
and the two are combined into one equation
\beqa
\sum_{\underline{i}}
\partial_{\underline{i}}(H\partial_{\underline{i}}y)=0.\label{eom3}
\eeqa
We note that 
the harmonic function 
gives nontrivial contribution to (\ref{eom3}), 
which means that 
the equations of motion of the D-4-brane worldvolume
are affected by the background D-8-brane.
By using (\ref{harm1}) and  $y>0$, 
the equation (\ref{eom3}) is written as  
\beqa
\sum_{\underline{i}}(\partial_{\underline{i}})^{2}
(y+\frac{c_{1}}{m})^{2}=0.
\eeqa
We choose the boundary condition 
\beqa
\left\{
  \begin{array}{lll} 
y&\rightarrow & y_{0} \ (>0)\\
V_{0}&\rightarrow & 0
  \end{array}
\right. \label{boundary1}
\eeqa
for $r \rightarrow \infty$, which means that
the D-4-brane lies at $y=y_{0}$. 
Then, the solution 
is obtained, with the following unusual form, 
as
\beqa
y(r)&=&[(\frac{c_{1}}{m}+y_{0})^{2}+\frac{c_{2}}{r^{2}}]^{1/2}
-\frac{c_{1}}{m}\nonumber\\
V_{0}&=&[(\frac{c_{1}}{m}+y_{0})^{2}+\frac{c_{2}}{r^{2}}]^{1/2}
-(\frac{c_{1}}{m}+y_{0})\label{sol1}
\eeqa
where $c_{2}$ is a constant proportional to the electric 
charge of the gauge field. 
The electric charge $Q_{1}$ is defined as \cite{gib1}
\beqa
Q_{1}=\int_{S_{3}} \star D \label{charge0} 
\eeqa
where $S_{n}$ is the n-sphere, $\star$ is the worldvolume Hodge
dual and $D$ is the 2-form defined by
$D^{ij}=-\frac{1}{T_{4}}
\frac{\delta {\cal L}_{D4}}{\delta F_{ij}}$.
Then, we have
\beqa
Q_{1}
=\int_{S_{3}} \star (HdV)=m c_{2}\Omega_{3}\label{charge1}
\eeqa
where $\Omega_{n}$ is the volume of the unit n-sphere. 
We note that in this definition
the string coupling $g_{s}=e^{\phi}$
is included in 
$Q_{1}$ since ${\cal L}_{D4}^{BI}$ is proportional to
the inverse of $g_{s}$. 

Since the solution has been
obtained explicitly,
we next give some pieces of
evidence that the solution corresponds to the fundamental
string ending on the D-4-brane. 
First, we discuss the massless limit of the solution (\ref{sol1}).
We assume here that the charge $Q_{1}$ is independent of $m$.
(The validity of the assumption is discussed later.)
Then, in the massless limit (\ref{masslesslim}),
the solution (\ref{sol1}) with 
(\ref{charge1}) behaves as
\beqa
y&\rightarrow & y_{0}+\frac{Q_{1}}{2\Omega_{3}r^{2}}\nonumber\\
V_{0}&\rightarrow & \frac{Q_{1}}{2\Omega_{3}r^{2}}.
\label{masslesssol}
\eeqa
The right hand side of (\ref{masslesssol}) 
is exactly the solution of 
a D-4-brane's worldvolume 
in the 10D {\it flat} spacetime, 
which corresponds to the fundamental string\cite{cal1}. 
So, it is expected to correspond to a fundamental string.
Next,
we check the preserved supersymmetry of the solution
by using ``superalgebras in brane backgrounds via brane
probes''\cite{sato1}\cite{sato2}\cite{sato3}.
The superalgebra in a D-8 brane background via a D-4-brane probe 
is given in ref. \cite{sato2} as
\beqa
\{ Q_{\alpha}^{+}, Q_{\beta}^{+}\}&=& 
2\int_{{\cal M}_{4}}d^{4}\xi\ \Pi_{\mu}
({\cal C}\Gamma^{\mu})_{\alpha\beta}
+2\int_{{\cal M}_{4}}d^{4}\xi{\cal P}^{(0)\underline{i}}
\partial_{\underline{i}}y
({\cal C}\Gamma_{y}\Gamma_{11})_{\alpha\beta}\nonumber\\
& &+\frac{2T_{4}}{4!}
\int_{{\cal M}_{4}}H^{5/4}dx^{\mu_{1}}..dx^{\mu_{4}}
({\cal C}\Gamma_{\mu_{1}..\mu_{4}}\Gamma_{11})_{\alpha\beta}
+2T_{4}\int_{{\cal M}_{4}}H^{5/4}dx^{\mu}dydV
({\cal C}\Gamma_{\mu y})_{\alpha\beta}\nonumber\\
& &+2T_{4}\int_{{\cal M}_{4}}
H^{5/4}(dV)^{2}({\cal C}\Gamma_{11})_{\alpha\beta}
+{\cal O}(\theta^{2})\label{spalgd4}
\eeqa
where $Q_{\alpha}^{+}= \frac{1+\Gamma_{\hat{y}}}{2}Q_{\alpha}$ is the
supercharge preserved in (\ref{d8sol})
and ${\cal M}_{p}$ is the worldspace of the D-p-brane.
${\cal P}^{(0)\underline{i}}$ is almost equivalent to
the conjugate momentum of 
$V_{\underline{i}}$. (The contributions of the Chern-Simons term
are subtracted.)
Substituting the solution for the right 
hand side of (\ref{spalgd4}),
the superalgebra can be written as
\beqa 
\{ Q_{\alpha}^{+}, Q_{\beta}^{+}\}&=&
4T_{4}\int_{{\cal M}_{4}}d^{4}\xi\
[ H^{1/4}(\frac{1+{\cal C}\Gamma_{\hat{1}\hat{2}\hat{3}\hat{4}}
\Gamma_{11}
}{2})_{\alpha\beta}
+H^{5/4}(\partial_{i}y)^{2}
(\frac{1+{\cal C}\Gamma_{\hat{y}}\Gamma_{11}}{2})_{\alpha\beta}]
\label{spalgd4'}.\ \ 
\eeqa
Since the three gamma matrix products $\Gamma_{\hat{y}},
{\cal C}\Gamma_{\hat{1}\hat{2}\hat{3}\hat{4}}\Gamma_{11}$ and 
${\cal C}\Gamma_{\hat{y}}\Gamma_{11}$ (arising in (\ref{spalgd4'}))
commute with each other,
all of them can be simultaneously diagonalized.
Since the square of each matrix product is
equal to the identity and each is traceless,   
both of
the matrices 
$\frac{1+{\cal C}\Gamma_{\hat{1}\hat{2}\hat{3}\hat{4}}\Gamma_{11}}
{2}$ 
and $ \frac{1+{\cal C}\Gamma_{\hat{y}}\Gamma_{11}}{2}$
are projection operators.
So, we conclude that the solution 
has 1/8 supersymmetry, hence 
is consistent with the target-space interpretation.
This also shows that the solution is a BPS state.

Now, we discuss the mass of the string.
Let us first consider the energy of the solution.
For this purpose,
we pass to the Hamiltonian formalism
as done in ref.\cite{cal1}.
If we assume that V is purely electric and that
only the scalar $y$ is excited,
$S_{D4}$ reduces to
\beqa
S_{D4}= -T_{4}\int d^{5}\xi
\sqrt{\{1-H(F_{0\underline{i}})^{2}\}
\{1+H(\partial_{\underline{i}}y)^{2}\}
+H^{2}(F_{0\underline{i}}\partial_{\underline{i}}y)^{2}
-H\dot{y}^{2}}
\eeqa
(where $H$ is the harmonic function).
The canonical momenta of $y$ and $V_{\underline{i}}$,
are defined respectively as
\beqa
P&=&\frac{T_{4}H\dot{y}}{\sqrt{\{1-H(F_{0\underline{i}})^{2}\}
\{1+H(\partial_{\underline{i}}y)^{2}\}
+H^{2}(F_{0\underline{i}}\partial_{\underline{i}}y)^{2}
-H\dot{y}^{2}}}
\nonumber\\
\Pi_{\underline{i}}&=&\frac{T_{4}H[F_{0\underline{i}}\{
1+H(\partial_{\underline{i}}y)^{2}\}-H\partial_{\underline{i}}y
(F_{0\underline{j}}\partial_{\underline{j}}y)
]}
{\sqrt{\{1-H(F_{0\underline{i}})^{2}\}
\{1+H(\partial_{\underline{i}}y)^{2}\}
+H^{2}(F_{0\underline{i}}\partial_{\underline{i}}y)^{2}
-H\dot{y}^{2}}}\ .
\eeqa
The Hamiltonian $\bar{H}$ is constructed as
$\bar{H}\equiv\int_{{\cal M}_{4}}d^{4}\xi {\cal H}$
where ${\cal H}$ is the Hamiltonian 
density given by
\beqa
{\cal H}=T_{4}
\sqrt{\{1+H(\partial_{\underline{i}}y)^{2}\}
(1+T_{4}^{-2}H^{-1}P^{2})+T_{4}^{-2}H^{-1}(\Pi_{\underline{i}})^{2}
+T_{4}^{-2}(\Pi_{\underline{i}}\partial_{\underline{i}}y)^{2}
}.\label{hami1}
\eeqa
We note that 
$\Pi_{\underline{i}}$ is subject to the constraint
$\partial_{\underline{i}}\Pi_{\underline{i}}=0$.
Substituting the solution (\ref{sol1})
for $\bar{H}$,
we can obtain the energy of the solution $E$
defined on the worldvolume.
We note that for a BPS solution like this case,
it generically happens that
the square root of $\bar{H}$ becomes a perfect square
and that the energy becomes a sum of 
the two parts:
the part originating from the D-p-brane and
that from the string.
so, we denote the first part of ${\cal H}$ as ${\cal H}_{1}$ 
and the second part as ${\cal H}_{2}$.
Concretely, the energy in this case takes the form
\beqa
E=T_{4}\int_{{\cal M}_{4}}d^{4}\xi[1+H(\partial_{\underline{i}}y)^{2}]
\equiv E_{1}+E_{2}.
\eeqa
The first term  $E_{1}$ 
is the ``energy'' of the D-4-brane itself,
and the second term $E_{2}$ 
is the energy of the excitation (i.e. the  string), 
both defined on the worldvolume.
Since we are interested in the second part,
we compute only $E_{2}$ here. 
$E_{2}$ is infinite in this case, but
if we regularize it by introducing a small parameter
$\delta$, 
we can get the energy for $r\geq \delta$ as
\beqa
E_{2}&=&T_{4}\int d\Omega_{3}
\int^{\infty}_{\delta} r^{3}dr (c_{1}+my(r))
(\partial_{\underline{i}} y(r) )^{2}\nonumber\\
&=&T_{4}Q_{1}(y(\delta)-y_{0}).
\eeqa
That is, the energy 
is (again) proportional to the difference of the coordinate.
Thus, we conclude that
{\it the energy defined on the brane's worldvolume
does not agree with the mass of the string 
in the case of D-branes in curved backgrounds.} 
We note that this result is rather reasonable, in a sense, 
in the case of $\tilde{g}_{00}\ne 1$,
because the energy has the same transformation property
as $\partial_{0}$ under the reparametrization of $\xi^{0}$.

Now, we construct the mass formula. 
Since 
it should be invariant 
under the reparametrization of $\xi^{0}$,
we propose the generalized mass formula $M$ for a string
described as a solution
of a D-p-brane's worldvolume,
as
\beqa
M
=\int_{{\cal M}_{p}}d^{p}\xi \sqrt{-\tilde{g}^{00}}{\cal H}_{2}
\label{stringmass}
\eeqa
where ${\cal H}_{2}$ is the second part of
the Hamiltonian density defined on the D-p-brane
(originating from the excitation corresponding to a string).

Let us calculate the mass of the string in this case,
based on the formula.
Substituting the solution for (\ref{stringmass}),
we find 
\beqa
M(\delta)&=& T_{4}\int_{{\cal M}_{4}}d^{4}\xi 
H^{5/4}(\partial_{\underline{i}}y)^{2}=
T_{4}\int d\Omega_{3}
\int^{\infty}_{\delta} r^{3}dr 
(c_{1}+my(r))^{5/4}(\partial_{\underline{i}} y )^{2}\nonumber\\
&=&\frac{4T_{4}Q_{1}}{5m}[\{(c_{1}+my_{0})^{2}+\frac{c_{2}}{\delta^{2}} 
\}^{5/8}
-(c_{1}+my_{0})^{5/4}].
\eeqa
On the other hand, 
the geodesic distance $l$ 
from the 
D-4-brane (lying at $y_{0}$) to the point parametrized by 
$y(\delta)$
is given by 
\beqa
l(y(\delta);y_{0})&\equiv& \int^{y(\delta)}_{y_{0}}
\sqrt{g_{yy}}dy
=\frac{4}{5}m^{1/4}[(y(\delta)+\frac{c_{1}}{m})^{5/4}
-(y_{0}+\frac{c_{1}}{m})^{5/4}].\label{geodesic}
\eeqa
So, we obtain the proportional relation:
\beqa
M=T_{4}Q_{1}\cdot l(y(\delta);y_{0}).\label{proporel}
\eeqa

Furthermore, 
we can show that 
the coefficient $ T_{4}Q_{1}$ reproduces
the tension of the fundamental string correctly.
To derive this, we discuss the unit electric 
charge for a (1,0) (i.e. a fundamental) string.
First, we review the discussion 
about the case of 
point charge solutions of the D-p-brane in the flat 
spacetime\cite{cal1} (and ref.\cite{sjrey2}).
Let us consider 
a triple junction of strings:
a (0,1) string, a (n,0) string and a (n,1) string.
If the string coupling 
$g_{s}$ is small,
it holds 
\beqa
\frac{\Delta T}{T_{(0,1)}}
=\frac{(g_{s})^{2}n^{2}}{2}\label{tensionrel1}
\eeqa
where $T_{(p,q)}$ is the tension of a (p,q) string and
$\Delta T$ is
the additional tension $\Delta T \equiv  T_{(n,1)}-T_{(0,1)}$.
On the other hand, 
the solution of a D-1-brane worldvolume
corresponding to 
the above string junction is 
given in ref.\cite{sjrey2}.
In the flat background with the ansatz
that $x^{0}=\xi^{0}, x^{1}=\xi^{1}, x^{9}=y(\xi^{1}), 
x^{m}=$ constant for m=2,..,8, and $\theta=0$,
the D-1-brane action is written as
\beqa
S_{D1}=-\frac{T_{1}}{g_{s}}\int d^{2}\xi
\sqrt{1+(\partial_{1}y)^{2}-(F_{01})^{2}
-(\partial_{0}y)^{2}}.
\eeqa
The solution of the D-1-brane's worldvolume as the triple string
junction with an electric charge $q_{1}$ is\cite{sjrey2}
\beqa
y(\xi_{1})=V_{0}(\xi_{1})=
\left\{
  \begin{array}{ll} 
-q_{1}\xi_{1}\ & {\rm for}\ \xi_{1}>0 \\
0 &{\rm for}\ \xi_{1}<0.
 \end{array}
\right. \label{triplesol}
\eeqa
The energy of the solution can be computed by using the Hamiltonian,
and the additional tension $\Delta T$ is also derived from this
correctly.
By taking into account the bending of the (n,1) string\cite{sjrey2},
it is obtained as
$\Delta T /T_{(0,1)}=(1/2)(F_{01})^{2}$.
Comparing this with (\ref{tensionrel1}),
the ``charge quantization condition'' 
$F_{01}=q_{1}=g_{s}n$ is deduced for a point charge $q_{1}$ (for
an integer $n$).
By T-dualizing with respect to 
the directions of $x^{m}$ for m=2,..,8,
the charge quantization condition
for a electric point charge $q_{1}$ of the D-p-brane's worldvolume
is shown to be
\beqa
q_{1}\equiv \frac{1}{(2\pi)^{p-1}}\int_{S_{p-1}} F_{0r}=g_{s}n.
\eeqa

Next, we discuss the case
which is related by T-duality to the case of (\ref{sol1}).
Let us suppose
a D-1-brane parallel to a subspace of
the worldvolume of a D-5-brane background, 
and that some number $n$ of fundamental strings 
are absorbed in the D-1-brane.
(This is also a BPS configuration
since 1/8 spacetime supersymmetry is preserved\cite{sen3}.)
If we consider a D-5-brane background solution,
%
the string coupling $g_{s}$ 
becomes a {\it local} function on 
the transverse coordinates $y^{a}$ ($g_{s}=e^{\phi}=H^{-1/2}$
where $H$ is the harmonic function on $y^{a}$), and
so is the tention $T_{(n,1)}$.
If the ``test'' D-1-brane is put near the D-5-brane, 
the string coupling $g_{s}$ is considered to be 
sufficiently small
around the D-1-brane.
So, the equation
(\ref{tensionrel1}) with $g_{s}=H^{-1/2}$
holds on the basis of the same discussion.
On the other hand,
suppose that a electric point charge $q_{1}'$ is added to
the D-1-brane's worldvolume with an excitation of a scalar field 
$y^{9}$.
Then, 
the D-1-brane action is
\beqa
S_{D1}
=-T_{1}\int d^{2}\xi \sqrt{1+H(\partial_{1}y^{9})^{2}-H(F_{01})^{2}
-H(\partial_{0}y^{9})^{2}}.\label{d1d51}
\eeqa
If we assume the existence of the correponding point charge 
solution,
the additional tension due to the field strength is
derived in the same way, as
\beqa
\frac{\Delta T}{T_{(0,1)}}=
\frac{1}{2(T_{1})^{2}}\cdot H^{-1}(\Pi_{1})^{2}
=\frac{1}{2}\cdot H (F_{01})^{2}.\label{tensionrel5}
\eeqa
So, comparing this with (\ref{tensionrel1})
we have
the charge quantization condition 
$H^{1/2}F_{01}=ng_{s}$ for an integer $n$.
By using the
the ``electric induction'' $D_{01}=-\frac{1}{T_{1}}\cdot 
\frac{\delta {\cal L}_{D1}}{\delta F_{01}}$ ($=HF_{01}$
in this case)
it can be rewritten in a more generic form as
\beqa
D_{01}=n.\label{chargeqcond3}
\eeqa
The higher dimensional D-brane cases are related to 
(\ref{chargeqcond3}) by T-dualities.\footnote{
We note that the background fields should also be transformed 
by T-dualities simultaneously.}
So, we have
\beqa
\frac{1}{(2\pi)^{p-1}}\int_{S_{p-1}}D_{0r}=n.\label{chargeqcond4}
\eeqa
This is the ``generalized''
charge quantization condition
for the point charge of the D-p-brane parallel
to the worldvolume of 
a D-(p+4)-brane background. 

Let us return to 
to the case of (\ref{sol1}). 
the left hand side of (\ref{chargeqcond4}) with $p=4$
is equivalent to
the charge $Q_{1}$ of (\ref{sol1}) multiplied by
$1/(2\pi)^{3}$.
So, 
the unit charge $Q_{1}$ for a (1,0) string in this case is 
$Q_{1}=(2\pi)^{3}$.
That is, 
the unit 
charge 
is the same as the one in the case of the flat background,
(This result is consistent from physical point of view,
since
the unit charge $Q_{1}$ 
is considered to be
independent of the background.) 
Thus, the proportional coefficient of (\ref{proporel}) for 
a (1,0) string is obtained as
$T_{4}Q_{1}=1/2\pi$, which is exactly
the tension of the string.
So, the quantity $M$ defined in (\ref{stringmass})
certainly gives the mass of the string correctly !

As another case, 
we consider the case of a test D-8-brane parallel to the background 
D-8-brane. 
The worldvolume action is given in (\ref{Dpaction}) for p=8.
In this case, we consider the ansatz
\beqa
\left\{
  \begin{array}{ll} 
    x^{i}=\xi^{i} 
      & (i=0,1,\cdots,8 )\nonumber\\
    y=y(r)(>0)& \nonumber\\
    V_{0}= V_{0}(r),& V_{\underline{i}}=0.
  \end{array}
\right.\label{pointsolansatz2}
\eeqa  
where $r$ is defined as 
$r\equiv\sqrt{\sum_{\underline{i}=1}^{8}
(\xi^{\underline{i}})^{2}}$.
Then, combined with $\theta=0$,
only the term 
including $\tilde{C}^{(9)}$ in $S_{D8}^{WZ}$
does contribute to the equations of motion 
($\tilde{C}^{(9)}_{01..8}=H^{\epsilon}$).
The expression of the induced worldvolume
metric $\tilde{g}_{ij}$ is the same as (\ref{wvmetric1})
except for the range of the indices 
(in this case $\underline{i},\underline{j}=1,2,..,8$).
Setting the same condition as (\ref{yveual1})
makes the determinant of $(\tilde{g}_{ij}+{\cal F}_{ij})$ 
simple, such as
${\rm det}(\tilde{g}_{ij}+{\cal F}_{ij})=H^{9\epsilon /2}$.
Then, 
we find the equations of motion 
\beqa
\sum_{\underline{i}}
\partial_{\underline{i}}(H^{-2\epsilon-2}\partial_{\underline{i}}y)
&=&0\\
\sum_{\underline{i}}
\partial_{\underline{i}}(H^{-\epsilon-1}\partial_{\underline{i}}y)
&=&0.
\eeqa
So, these two equations are again compatible only if
$\epsilon=-1$, and 
the equations to solve become a single equation 
\beqa
\sum_{\underline{i}}(\partial_{\underline{i}})^{2}y=0.\label{eom'3}
\eeqa
We note that unlike the D-4-brane case,
the harmonic function $H$ does not appear in  
(\ref{eom'3}). 
So, choosing the same boundary condition as (\ref{boundary1}),
the solution is obtained easily as
\beqa
y&=&y_{0}+\frac{c_{2}'}{r^{6}}\nonumber\\
V_{0}&=&\frac{c_{2}'}{r^{6}}\label{sol2}
\eeqa
where $c_{2}'$ is a constant proportional to the electric
charge of the solution.
By using the definition of the charge
similar to (\ref{charge0}),
we have the electric charge $Q_{1}'$:
\beqa
Q_{1}'\equiv -\frac{1}{T_{8}}
\int_{S_{7}}\star(\frac{\delta {\cal L}_{D4}}{\delta F_{ij}})
=\int_{S_{7}}\star(dV)
=6c_{2}'\Omega_{7}.
\eeqa
We note that the form
of this point charge 
solution (\ref{sol2}) 
is completely the same as 
that 
of the D-8-brane worldvolume
in the {\it flat} background.

Here, 
We derive the preserved supersymmetry of the solution
(\ref{sol2}) in the same way.
1/4 supersymmetry is expected to be preserved.
The superalgebra in a D-8 brane background via a D-8-brane probe 
is\cite{sato2}
\beqa
\{ Q_{\alpha}^{+}, Q_{\beta}^{+}\}=
2\int_{{\cal M}_{8}}d^{8}\xi\ \Pi_{\mu}
({\cal C}\Gamma^{\mu})_{\alpha\beta}
+2\int_{{\cal M}_{8}}d^{8}\xi{\cal P}^{(0)\underline{i}}
\partial_{\underline{i}}y
({\cal C}\Gamma_{y}\Gamma_{11})_{\alpha\beta}\nonumber\\
\ \ \ \ \ \ \ \ \ \ \ \ \nonumber\\
+\frac{2T_{8}}{5!}\int_{{\cal M}_{8}}
H^{5/4}dx^{\mu_{1}}..dx^{\mu_{5}}dydV
({\cal C}\Gamma_{\mu_{1}..\mu_{5}y})_{\alpha\beta} 
+\frac{2T_{8}}{4!}\int_{{\cal M}_{8}}
H^{5/4}dx^{\mu_{1}}..dx^{\mu_{4}}(dV)^{2}
({\cal C}\Gamma_{\mu_{1}..\mu_{4}}\Gamma_{11})_{\alpha
\beta} \nonumber\\
+ 2T_{8}\int_{{\cal M}_{8}}H^{5/4}dx^{\mu}dy(dV)^{3}
({\cal C}\Gamma_{\mu y})_{\alpha\beta}
+2T_{8}\int_{{\cal M}_{8}}
H^{5/4}(dV)^{4}({\cal C}\Gamma_{11})_{\alpha\beta}
+{\cal O}(\theta^{2}).
\ \ \ \ \ \ \ \ \ \ \ \ \ \ \ \ \ 
\label{spalgd8}
\eeqa
The momentum $ \Pi_{\mu} $
includes the following two terms: 
\beqa
\Pi_{\mu}=\Pi_{\mu}^{(0)} + \frac{T_{8}}{8!}H^{-2}my
\epsilon^{0i_{1}..i_{8}}\partial_{i_{1}}x^{\nu_{1}}..
\partial_{i_{8}}x^{\nu_{8}}\epsilon_{\mu\nu_{1}..\nu_{8}y}
\eeqa
where $\Pi_{\mu}^{(0)}$ is the contribution of $S^{BI}_{D8}$
and the second term is that of $S^{WZ}_{D8}$ 
(where $\epsilon^{01..8}=1$).
Substituting the solution for the right hand side of (\ref{spalgd8}),
the superalgebra can be written as
\beqa 
\{ Q_{\alpha}^{+}, Q_{\beta}^{+}\}&=&
4T_{8}\int_{{\cal M}_{8}}d^{8}\xi\
H^{1/4}(\partial_{\underline{i}}y)^{2}
(\frac{1+{\cal C}\Gamma_{\hat{y}}\Gamma_{11}}{2})_{\alpha\beta}
\label{spalgd8'}.
\eeqa
By the same discussion as 
done in the D-4-brane case, 
it is shown
that 1/4 supersymmetry is preserved in this configuration,
which is consistent with the spacetime interpretation.
So, we interpret the solution (\ref{sol2}) as a fundamental string 
again.

Applying the mass formula (\ref{stringmass})
to this case,
we obtain the result:
\beqa
M &\equiv& \int_{{\cal M}_{8}}\sqrt{-\tilde{g}^{00}}{\cal H}_{2}
=T_{8}\int_{{\cal M}_{8}}d^{8}\xi H^{1/4}
(\partial_{\underline{i}} y )^{2}\nonumber\\
&=&T_{8}\int d\Omega_{7}
\int^{\infty}_{\delta} r^{7}dr(c_{1}+my(r))^{1/4}
(\partial_{\underline{i}} y )^{2}\nonumber\\
&=& T_{8}Q_{1}'\cdot l(y(\delta);y_{0}).
\eeqa
Based on the same discussion as that done
in the D-4-brane case, the unit electric charge 
$Q_{1}'$ for a (1,0) string ending on a D-8-brane can also be 
derived as
$Q_{1}'= (2\pi)^{7}$. 
So,
the tension of the string is reproduced correctly again,
and
(\ref{stringmass}) also gives
the mass of the string for the solution (\ref{sol2}) correctly.
Therefore, 
we conclude that (\ref{sol2}) is the correctly 
generalized mass formula, 
which holds when the background of the D-brane is curved. 
We note that 
the energy defined on the worldvolume again 
gives the difference of the coordinate $y$ multiplied by
$T_{8}Q_{1}'$.
That is, this case is another example that
the worldvolume energy does not agree with the mass of the
string.

\section{Summary and discussions}

In summary, 
we have proposed (\ref{sol2}) as the generalized mass formula 
which holds when the background of the D-brane is curved,
and have
shown explicitly using the two examples
that the formula certainly gives
the mass of a fundamental string
described as a BPS solution of a D-brane's worldvolume.
In addition, based on the obtained formula,
we can see that
the mass of the string agrees with the worldvolume energy
only in the cases $\tilde{g}_{00}=-1$ 
(where
$\tilde{g}_{ij}$ is the induced worldvolume metric).
which include
the case discussed in ref.\cite{cal1}.

Here, we
discuss 
the consistency of the mass formula from another point of view,
especially
focusing on the factor $\sqrt{-\tilde{g}^{00}}$.
Suppose we consider the D-4-brane embedded parallel to the
D-8-brane background (\ref{d8sol}) {\it with no excitation of
worldvolume fields} (i.e. $x^{i}=\xi^{i}$ for i=0,..,4 and
$x^{5},..,x^{8},y$ : constants).
Then, on the analogy of the mass formula of a (p,q) string
given by Sen
in ref.\cite{sen1},
the target-space mass $m_{D4}$ of the D-4-brane 
should be proportional to its spatial volume element 
{\it measured by the geodesic distances} in the spacetime.
So, it should be given by
\beqa
m_{D4}=\int_{{\cal M}_{4}}d^{4}\xi T_{4} e^{-\phi}
\sqrt{{\rm det}\tilde{g}_{\underline{i}\underline{j}}'}
\eeqa 
where $\tilde{g}_{\underline{i}\underline{j}}'$ 
is the induced {\it world-space} metric
of the D-4-brane. 
In fact, the mass $m_{D4}$ 
obtained in this way can be shown to agree with
the quantity $M_{D4}$ defined as
\beqa
M_{D4}
=\int_{{\cal M}_{4}}d^{4}\xi \sqrt{-\tilde{g}^{00}}{\cal H}_{1}
(=\int_{{\cal M}_{4}}d^{4}\xi H^{1/4}\cdot 1)
\eeqa
where ${\cal H}_{1}$ is the first part (i.e. 
originating from the D-4-brane) of
the Hamiltonian density of the solution (\ref{sol1})
defined on the D-4-brane.
This means that the information of the D-4-brane mass
can also be extracted from the solution (\ref{sol1})
by integrating the Hamiltonian density
multiplied by the factor $\sqrt{-\tilde{g}^{00}}$
with respect to world-space coordinates.
Thus, the factor $\sqrt{-\tilde{g}^{00}}$
arising in the formula is consistent in this sense, too.

Finally, let us discuss 
the worldvolume interpretation of the mass formula (\ref{stringmass}).
If we define a worldvolume proper time $\tau$ as 
$d\tau\equiv \sqrt{-\tilde{g}_{00}}d\xi^{0}$,
the Hamiltonian density multiplied by $\sqrt{-\tilde{g}^{00}}$
might be regarded as the energy density defined
with respect to $\tau$.
So, we might say that from worldvolume point
of view,
the mass of the string is interpreted as
``the energy defined 
with respect to the worldvolume proper time''.

\parbigskipn

{\Large\bf Acknowledgement}

\parbigskipn
I would like to thank Taro Tani for useful discussions
and encouragement. 
I would also like to thank Y. Imamura, 
Tsunehide Kuroki and Shinya Tamura
for useful discussions
in computing the energy of the solution.

\parbigskipn

\end{document}